\documentclass[epjst]{svjour}

\usepackage[numbers,compress]{natbib}
\usepackage{graphics}
\usepackage[utf8]{inputenc}
\usepackage[T1]{fontenc}
\usepackage{amsmath}
\usepackage{amssymb}
\usepackage{hyperref}
\usepackage{changes}
\usepackage{hyperref}
\usepackage{fancyvrb}
\usepackage{siunitx}
\usepackage{changes}

\DeclareSIUnit\Wh{Wh}
	
\usepackage{lineno}

\begin{document}
\title{Energy-efficient recurrence quantification analysis}
\author{Norbert Marwan\inst{1,2,3}\fnmsep\thanks{\email{marwan@pik-potsdam.de}}}
\institute{Potsdam Institute for Climate Impact Research (PIK), Member of the  Leibniz Association,
Telegrafenberg A31, 14473 Potsdam, Germany \and
University of Potsdam, Institute of Geoscience, Karl-Liebknecht-Straße 32, 14476 Potsdam, Germany \and
University of Potsdam, Institute of Physics and Astronomy, Karl-Liebknecht-Straße 32, 14476 Potsdam, Germany
}
%

%
\abstract{
Recurrence quantification analysis (RQA) is a widely used tool for studying 
complex dynamical systems, but its standard implementation requires 
computationally expensive calculations of recurrence plots (RPs) and line
length histograms. 
This study introduces strategies to compute RQA measures directly from 
time series or phase space vectors, avoiding the need to construct 
RPs. The calculations can be further accelerated and optimised by applying a random 
sampling procedure, in which only a subset of line structures is 
evaluated. These modifications result in shorter run times, less memory 
use and access, and lower overall energy consumption during analysis while maintaining accuracy. This makes them especially appealing for large-scale data analysis and 
machine learning applications.
The ideas are not limited to diagonal line measures, but can likewise 
be applied to vertical line-based measures and to recurrence network 
measures. By lowering computational costs, the 
proposed strategies contribute to energy saving and sustainable data analysis, and 
broaden the applicability of recurrence-based methods in modern research contexts.
} 
\maketitle
\section{Introduction}
\label{sec_introduction}

In the light of sustainable energy production and climate change, 
energy efficiency in modelling and data analysis is an important and growing topic \cite{zhang2020geomoddev,paul2023}. 
As the demand for data processing grows, it is crucial to focus on reducing related
energy consumption \cite{murana2019,kocot2023,bruers2024}. This problem is increasingly attracting attention
across various fields, but it remains in its early stages. Among the many goals and 
actions that help address energy-efficient computations \cite{bruers2024}
and are especially feasible in daily work and short-term contexts, the key measures
include educating about this topic and enhancing implementations \cite{augier2021}.

In the context of sustainable data analysis, energy efficiency is becoming 
increasingly critical. This is particularly relevant for computationally intensive 
methods such as recurrence quantification analysis (RQA), a powerful tool in 
nonlinear data analysis \cite{eckmann87,marwan2007,webber2015,zbilut92}. This 
framework has shown high potential in investigating 
diverse research questions across many disciplines \cite{marwan2008epjst}
and its integration with machine learning (ML) has recently sparked considerable interest \cite{marwan2025}.

However, this potential comes at a cost: RQA stands out as a particularly 
energy-intensive method in the ML toolkit. Unlike linear 
techniques, RQA relies on pairwise comparisons in high-dimensional data, 
resulting in computational complexity that scales quadratically with data size. 
In ML workflows, this challenge is amplified: RQA is often embedded in iterative 
processes -- such as feature extraction, hyperparameter tuning, or real-time 
inference -- where its energy demands compound with those of the broader pipeline. 
For example, extracting RQA-based features from thousands of time series segments or 
combining RQA with neural networks in hybrid models can result in high energy 
consumption costs \cite{rozycki2025,poenaruolaru2025}. Moreover, the nonlinear nature of RQA resists simple optimisations,
making efficient implementations not just a performance concern, but a necessity 
for sustainable ML. Addressing this gap is critical to unlocking RQA's potential 
in data-intensive fields like climate science, biomedicine, and IoT --
where energy-efficient algorithms are crucial for scalable, real-world applications \cite{rozycki2025}.
Beyond reducing energy consumption, optimised algorithms generally provide faster
computations, enabling real-time applications and large-scale deployments \cite{rawald2015,zhang2020geomoddev,poenaruolaru2025}. 

The standard RQA algorithm begins with calculating the RP, which has a 
computational complexity of $\mathcal{O}(N^2)$ \cite{marwan2007}. The lines 
formed by consecutive points in the RP are then detected, and their lengths
are measured. As we consider diagonal and vertical lines, the computational 
complexity for each of these steps is also of order $\mathcal{O}(N^2)$
(also in the case of symmetric RPs, where we would consider only one triangle
of the RP). This complexity requires many calculation steps and, therefore,
also increases computation times.
In the last decade, various approaches have been proposed to speed up RQA 
calculations. Classical approaches utilise 
parallelisation and use multiple graphics processing unit (GPU) devices
\cite{rawald2015,rawald2017}. The latter distributes the calculations 
of RPs and line lengths on different GPU devices using the divide and recombine
paradigm \cite{rawald2014}. Although the acceleration of the computations
is remarkable, the complexity of the calculations has not changed.
An approach that can indeed reduce the computational complexity extremely
is a numerical and geometrical approximative ansatz, which can heavily accelerate the 
calculations by several magnitudes, but with
the costs of increasing uncertainties \cite{schultz2015,spiegel2016}.
More recently, the introduction of microstate recurrence analysis \cite{corso2018}
has offered another (random sampling-based) approximative but fast approach, 
with much less uncertainty
in the RQA measures \cite{daCruz2025,ferreira2025}. Here, the line length distributions, required
for the RQA measures, are estimated from randomly sampled sub-matrices from the
RP.

In the following, we will consider the original approach of RQA calculations,
without parallelisation (although it would be possible) and without
numerical/ geometrical approximation,
but using the diagonal line structures of interest and developing some
optimisation strategies. This optimisation follows the action numbers
M2 (``Reduce and compress data having the anticipated scientific 
value of the retained information and the resource requirements in mind'') and
M7 (``Design software for optimized energy consumption and provide tools to 
measure it'') \cite{bruers2024}. 
These actions are part of the ``Call-to-action in digital transformation'', which outline concrete measures developed by members of the ErUM-Data community during a dedicated workshop, aiming to reduce greenhouse gas emissions and promote the use of renewable energy within data-intensive research \cite{bruers2024}. For the sake of
simplicity, only the diagonal lines are considered here; the same approach
can be applied for calculating the measures based on the vertical lines
or even more complex patterns, such as triangle motifs (corresponding to
the recurrence configuration that two neighbours of a state vector 
are also neighbours, which is required for recurrence network measures \cite{marwan2009b,donner2010b}).
The first optimisation will be very simple and has surely already been applied 
in several implementations. The second one is inspired by the microstates
approach by \citet{daCruz2025}. These approaches are compared to a 
simple, straightforward standard implementation without any optimisations, 
as someone lacking knowledge of algorithmic and language-specific 
optimisations (such as those available in Julia) would perform it.
The algorithms are implemented using the Julia language because it 
enables easy measurement of the performance of the implementations.

\section{Recurrence Plots and Recurrence Quantification Analysis}
\label{sec_recurrence}

A recurrence plot (RP) is a square matrix $\mathbf{R}$ indicating the similarity
of two states $\vec{x}(i)$ and $\vec{x}(j)$ at different points
in time $i$ and $j$:
\begin{equation}\label{eq_RP}
R(i,j) = \Theta\left(\varepsilon - \|\vec{x}(i) - \vec{x}(j)\|\right),
\end{equation}
with states $\vec{x} \in \mathbb{R}^d$ ($d$ the dimension of the
phase space), $i, j = 1, \ldots, N$ the time indices, $\Theta$ the 
Heaviside function, and $\varepsilon$ the recurrence threshold \cite{marwan2007}.
The RP exhibits typical large-scale appearances, depending on the 
system's dynamics \cite{eckmann87,marwan2007}. It further contains small-scale line structures
that represent temporally close evolution of the phase space trajectory
pairs (diagonal lines) or trapped states (vertical lines). The 
distributions of these line lengths provide insights into the 
dynamics of the system and are used to define measures of complexity within 
recurrence quantification analysis (RQA).

A typical RQA measure is the determinism measure (DET), which
quantifies the ratio of recurrence points ($R(i,j) = 1$)
which form diagonal lines and the total number of recurrence points, 
and is based on the histogram $P(\ell)$ of the
occurrences of lines of length $\ell$,
\begin{equation}
DET = \frac{\sum_{\ell=2}^N \ell P(\ell)}{\sum_{\ell=1}^N \ell P(\ell)}.
\end{equation}
This measure is related to predictability, with large values for
highly predictable dynamics and low values for non-predictable dynamics.
It can be used, e.g., to identify regime changes or classify different
regimes or systems \cite{colafranceschi2007,garciaochoa2009,marwan2015,mosdorf2012}.

For a more general overview of the method and its application potential, the reader is referred to \cite{marwan2008epjst,marwan2023}.

\section{Optimisation Strategies}
\label{sec_optimisation}

The following implementations have been tested on a single core
of the high-performance cluster ``Eunice Newton Foote''
at the Potsdam Institute of Climate Impact Research
(AMD EPYC 9554 processors with scalar frequencies of up to 3.75 GHz 
and 6 GByte DDR5 memory per core). The performance of an implementation
was tested using the Julia macros \texttt{@time} and \texttt{@timed}. 
The calculations were performed 100 times to ensure reliable 
statistical estimates of the performance measurements.

As test data, the
three components of the R\"ossler system (using standard parameters
$a=1.2$, $b=0.2$, and $c=5.7$)
are used, with length 25,000 and sampling time $\Delta t = 0.2$ \cite{roessler1976}.
The R\"ossler system was integrated using a Tsitouras 5/4 Runge–Kutta–Solver
as implemented by the DifferentialEquations.jl package in Julia.

Recurrence, Eq.~(\ref{eq_RP}), is calculated for all three components
of the R\"ossler system, using a threshold $\varepsilon$ of 10\% of
the range of the values of all three components, i.e.,
$\varepsilon = 0.1(\max(\vec{x}) - \min(\vec{x}))$.

\subsection{Histogram Estimation without RP (RQA\_woRP)}
\label{sec_woRP}

\begin{figure}[tbp]
\begin{center}
   \includegraphics[width=\columnwidth]{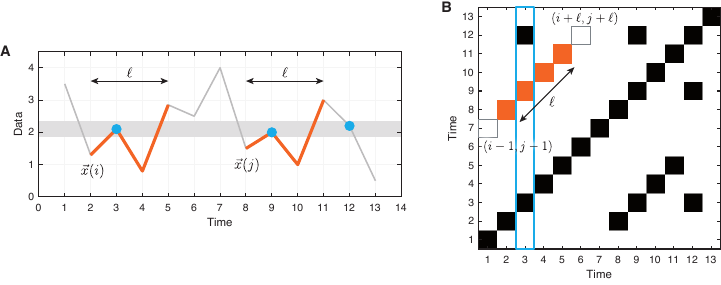}
   \caption{(A) Time series sequence and (B) corresponding recurrence plot. The sequence
   at time points 2 to 5 (orange) repeats four times within a small error $\varepsilon$ (here $\varepsilon = 0.25$) 
   during the interval 8 to 11, forming a diagonal
   line with points $(i=2,j=8)$, $(i+1=3,j+1=9)$, $(i+2=4,j+2=10)$, 
   and $(i+3=5,j+3=11)$ in the RP (orange points). The line is preceded by condition
   Eq.~(\ref{eq_start_condition}), $\|\vec{x}(i-1) - \vec{x}(j-1)\| > \varepsilon$,
   and followed by condition Eq.~(\ref{eq_end_condition}), 
   $\|\vec{x}(i+\ell) - \vec{x}(j+\ell)\| > \varepsilon$ ,
   (grey boxes in panel (B)), ensuring the lines start and end points.
   The length of the
   diagonal line ($\ell = 4$) can be measured in the RP or directly from the time series.
   The state at time $3$ recurs (within the $\varepsilon$ uncertainty, 
   grey horizontal bar in panel (A)) at time $9$ and $12$ (blue dots), indicated by the
   points in the column $3$ (marked by blue box).
   }\label{fig_diagonalline}
\end{center}
\end{figure}

A line structure in an RP is a sequence of pairs of time indices.
A diagonal line in the RP of length $\ell$ with the coordinates
$\{(i, j), (i+1, j+1), \ldots, (i+\ell-1, j+\ell-1)\}$
represents that the sequences of states 
$\{\vec{x}(i),\vec{x}(i+1),\ldots,\vec{x}(i+\ell-1)\}$ and
$\{\vec{x}(j),\vec{x}(j+1),\ldots,\vec{x}(j+\ell-1)\}$
are similar within the recurrence uncertainty defined by the
threshold $\varepsilon$ (Fig.~\ref{fig_diagonalline}).

It is clear that we can find line lengths in the RP without the
RP. We can test the condition 
\begin{equation}\label{eq_rec_condition}
\|\vec{x}(i) - \vec{x}(j)\| \le \varepsilon
\end{equation}
directly at the series of the phase space vectors (or the time series)
for increasing indices $i$ and $j$. Considering the condition
\begin{equation}\label{eq_start_condition}
\|\vec{x}(i-1) - \vec{x}(j-1)\| > \varepsilon
\end{equation}
for the start of a line and 
\begin{equation}\label{eq_end_condition}
\|\vec{x}(i+\ell) - \vec{x}(j+\ell)\| > \varepsilon
\end{equation}
as the end of a line, we can get the distribution of line lengths
$P(\ell)$. This is a very simple and straightforward calculation
of $P(\ell)$ and allows us to get the RQA measures without previous
calculating the RP beforehand. The resulting histogram of the 
diagonal lines will be exactly the same as the one obtained from
the RP. It will not reduce the numerical complexity
but the number of calculation steps, and should, therefore,
result in more efficient and faster RQA calculations when
the RP is not required. It also reduces the number of memory
allocations, as we do not need to store and calculate the matrix $\mathbf{R}$.

This kind of optimising the histogram calculation without
calculating the RP beforehand is surely not novel. It has been used
by the author before \cite{marwan_rqaopenmp} and also
have been implemented by others (e.g., \cite{martinovic2019}). For example, such
implementation is available in \cite{marwan_rqaopenmp} and was used in the
run-time comparisons in the studies by \citet{rawald2014,rawald2017} and
\citet{spiegel2016}.

The comparison with the standard implementation using RP-based
line histogram calculation shows a speed-up by a factor
of more than 2 (Tab.~\ref{tab_generalperformance}). The number
of memory allocations dropped from 7 to 3. The just-in-time compiler
(JIT) in Julia allows some internal optimisation steps, e.g.,
by applying the macro \texttt{@inbounds} to the for-loops \cite{sherrington2024}. Using
it can further accelerate the implementation. Here, it
finally speeds up the implementation by a factor of more than 4. In the following
analysis, we will use the \texttt{@inbounds} optimised version.

\begin{table}[htbp]
\caption{General performance values of the different implementations when
calculating the line length histograms required for RQA and using the R\"ossler
system with all three components and of length $N=25,000$; number of samplings
for the sampling version and microstates $M=4N = 100,000$, except for (RQA\_Samp$^2$);\\
$^1$ the calculation directly from the time series (RQA\_woRP) has been performed
additionally with the Julia instruction \texttt{@inbounds} which can accelerate
the code further\\
$^2$calculation of RQA\_Samp using a smaller number of samplings $M=0.2N = 5,000$.}
\begin{center}
\begin{tabular}{lrrrr}
Implementation					&Computation time	&Speedup	&Allocations	&Energy\\
\hline
RQA from RP	(RQA\_RP)			&\SI{2.01}{\s}	&1		&7		&\SI{27.9}{\milli\Wh}\\
RQA without RP (RQA\_woRP)		&\SI{0.88}{\s}	&2.3		&3		&\SI{12.2}{\milli\Wh}\\
RQA without RP$^1$ (RQA\_woRP$^1$)	&\SI{0.48}{\s}	&4.2	&3	&\SI{6.7}{\milli\Wh}\\
RQA from sampling (RQA\_Samp)	&\SI{0.25}{\s}	&8.0		&3		&\SI{3.5}{\milli\Wh}\\
RQA from sampling$^2$ (RQA\_Samp$^2$)	&\SI{0.012}{\s} &168	&3	&\SI{0.17}{\milli\Wh}\\
RQA from microstates			&\SI{0.017}{\s}	&1148	&29		&\SI{0.24}{\milli\Wh}\\
\hline
\end{tabular}
\end{center}
\label{tab_generalperformance}
\end{table}%

\subsection{Sampling-based Histogram Estimation (RQA\_Samp)}

Based on the histogram estimation directly from the data, and without
RP, we can further reduce the calculation costs by applying a 
random sampling scheme. Instead of sequentially testing every
index $j$ from 1 to $N$ for every $i$, we can randomly draw indices
$i$ and $j$ and test whether this pair fulfils the conditions given by 
Eqs.~(\ref{eq_rec_condition}) and (\ref{eq_start_condition}),
i.e., correspond to the beginning of a diagonal line. If not, we randomly draw
a new pair $i$ and $j$. If yes, we can measure the line length
as is done in RQA\_woRP, but stop after the end of a line and randomly
sample a new line. By this sampling schema, we get only a subset
of lines in the histogram $P(\ell)$, depending on the number
$M$ of samplings. Using a smaller number of $M$, e.g., 
$M = 4 N$, we can significantly reduce the number of
calculations to estimate $P(\ell)$. The resulting histogram $P(\ell)$
is a direct approximation of the histogram as it would be obtained using
the RP\_woRP algorithm described in Subsect.~\ref{sec_woRP}. 
Importantly, this approximation concerns the same underlying distribution, and the standard RQA measures can be computed from $P(\ell)$ without any modification.
Even for small values of $M$, this distributional approximation is 
already quite accurate. 

In general, this sampling approach is, somehow,
similar to estimating RQA measures using randomly sampled 
microstates as explained in \cite{daCruz2025}, but with the difference
that RQA measures (such as DET) can be directly estimated from the
histogram $P(\ell)$, i.e., they are not reconstructed from a different
(the microstated) distribution, and, thus, will 
result in higher accuracy when using the same sampling size $M$.

Comparing to the standard estimation algorithm (RQA\_RP) and the 
direct one (RP\_woRP), we find a further speed-up depending
on the number $M$ (Tab.~\ref{tab_generalperformance}). 
The memory allocations are the same as for the direct histogram
estimation without RPs. A large number of samples $M = 4N$,
the computation time is halved; further reducing $M$ to 20\% of
the data length
speeds up the RQA estimation by two orders of magnitude.

\begin{figure}[htbp]
\begin{center}
   \includegraphics[width=.8\columnwidth]{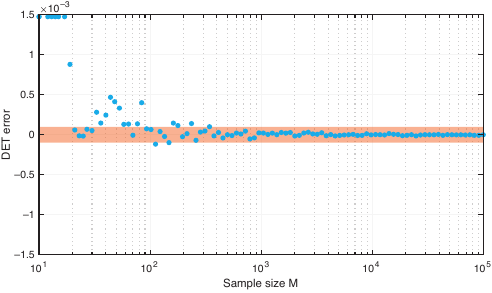}
   \caption{Estimation error $DET_\text{sampled} - DET_\text{true}$ derived from sampled line length
   histogram for the R\"ossler system with 25,000 data points
   and for sampling size ranging from $M=10$ to $100,000$. 
   The orange region indicates the error range of $\pm 10^{-4}$.
   A random sampling of only 500 line segments already results in minor errors
   smaller than $10^{-4}$.
   }\label{fig_samplesize}
\end{center}
\end{figure}

Therefore, it might be desired to have a small $M$, but it obviously
causes worse estimates of the line length measures.
Considering the error $DET_\text{sampled} - DET_\text{true}$
shows that increasing $M$ reduces the error of the DET estimation when obtained with this
sampling schema (Fig.~\ref{fig_samplesize}). For the considered data
with $N=25,000$, a sampling size of $M=500$ would already result in excellent
estimations of DET with errors of $<10^{-4}$, using $M=5,000$ the
estimation errors drop below $10^{-5}$.

\subsection{Efficiency by computation time}

\begin{figure}[htbp]
\begin{center}
   \includegraphics[width=.7\columnwidth]{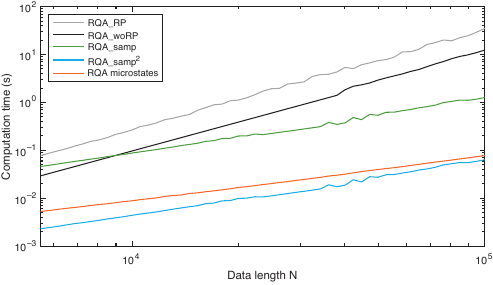}
   \caption{Computation times for the RQA implementations using
   RP (RQA\_RP), without RP (RP\_woRP), with a sampling scheme
   (RP\_Samp), and using a recurrence
   microstates approximation for increasing data length $N$ (R\"ossler 
   system, same parameters as in Fig.~\ref{fig_samplesize}). 
   The sampling size $M$ for RP\_Samp and microstates is $M=4N = 100,000$,
   and for RP\_Samp$^2$ it is $M=0.2N = 5,000$.
   }\label{fig_datasize}
\end{center}
\end{figure}

One aspect of efficient implementations is the reduced number of 
calculation steps, saving energy and also computation time. The first
approach of calculating the RQA directly from the data, without the
RP (RP\_woRP) shows already a significant reduction of computation time of more
than 50\%, because
calculating the RP is not required (Fig.~\ref{fig_datasize}). Further optimisation using
the sampling schema (RP\_Samp) reduces the computation time remarkably.
This reduction depends on the sampling size $M$. However, it is not
necessary to scale the sampling size $M$ with the squared size $N$ of the
data. We find high accuracy already when we scale $M$ linearly with the
size $N$, e.g., by $M=4N$ or $M=0.2N$, as it is used here in the comparison.
This results in a lower scaling exponent of the computation time
as data length increases (Fig.~\ref{fig_datasize}). This means, the
longer the data, the more efficient this method becomes.

We also note that for small data ($N<10,000$), the sampling schema
is not necessarily faster than RQA\_woRP.
Additionally, uncertainties in the RQA results increase for 
shorter time series when using 
sampling-based approaches, affecting both RQA\_Samp and the microstates 
method (see appendix, Fig.~\ref{fig_datasize_error}), suggesting that 
RQA\_woRP is preferable in such cases.

Using the same number $M$ in the recurrence microstates approach \cite{ferreira2025},
the computation time is, in general, smaller (although
it might not be directly comparable, because the microstates
calculation is based on compiled C code highly optimised for Julia \cite{microstates_software}), 
but with the cost of less 
accurate results (see
appendix, Figs.~\ref{fig_microstates_samplesize} and \ref{fig_datasize_error}).

The reduced computation time directly translates into lower energy consumption. 
Estimating a specific power draw per CPU core for RQA calculations is not
easy, as it depends on the workloads, architecture, computer system 
(e.g., HPC, laptop), underlying operating systems, and other factors 
\cite{kistowski2016, kocot2023,murana2019}. 
However, assuming an average power draw of $\approx$\SI{50}{\watt} per
CPU core as a realistic
baseline, replacing 
RQA\_RP (\SI{2.06}{\second} for data with length $N=25,000$) with 
RQA\_Samp (\SI{0.26}{\second}) reduces the energy usage from approximately 
\qtyrange{0.029}{0.0036}{\Wh} per run, corresponding to an energy 
saving of about \SI{0.025}{\Wh},
or almost 90\%. Using the even faster RQA\_Samp$^2$ (\SI{0.013}{\second}) 
further lowers the consumption to \SI{0.00018}{\Wh}, i.e., a saving of 
roughly \SI{0.028}{\Wh} per run. While these savings are small for a 
single computation, they accumulate substantially when processing large data sets, 
performing repeated analyses, or running parallel workflows on HPC systems. 
For example, 100,000 runs of a typical RQA 
of data with length $N=25,000$ accumulate to an energy saving of \SI{2.8}{\kWh},
equivalent to $\approx$\SI{1.0}{\kilo\gram} of CO$_2$
emissions (assuming \SI{363}{\gram} CO$_2$ per \unit{\kWh} based on the
German electricity mix in 2024 \cite{uba2025}).

\section{Discussion and conclusion}

Efficiency gains in data analysis can be achieved at multiple levels. 
Low-level optimisations, such as compiler hints (e.g., \texttt{@inbounds} in 
Julia's JIT compiler) or vectorisation directives, can reduce 
execution time without changing the underlying algorithm. However, 
more substantial improvements are typically obtained through 
algorithmic modifications, such as avoiding intermediate data 
structures and calculation steps, or applying sampling schemes, like the RQA\_woRP
and RQA\_Samp approaches.

These discussed approaches demonstrate that recurrence-based 
complexity measures can be obtained in a direct and computationally 
efficient way without the need for pre-computed RP.
When this idea is combined with a sampling strategy, in which only 
a fraction of the possible structures are analysed, the computational 
costs can be reduced even further. This leads not only to faster 
execution, but also to lower memory requirements and energy 
consumption, which are relevant considerations in modern large-scale 
data analysis pipelines.

Such efficiency gains are particularly important in applications where 
RQA is employed in combination with machine learning techniques, 
as efficient computation allows more extensive training and validation 
procedures and facilitates the systematic exploration of the model 
architectures. They are equally relevant in the context of large-scale (big)
data analysis, where reduced memory and runtime requirements make it possible
to apply RQA to very long time series, large ensembles of signals, or
even continuous data streams. By strengthening both domains, the proposed
approach enhances the accessibility and practical utility of RQA in 
modern data-driven research.
Furthermore, the reduced computational overhead may facilitate the integration of RQA methods into real-time or embedded systems, where resources are limited.

The choice of which approach to use depends on the research questions,
data characteristics, and computational constraints. 
For small datasets ($N < 10,000$) and when RPs are not needed, the RQA\_woRP 
approach is the best choice, ensuring high accuracy and 
reasonable computation time. 
RQA\_Samp becomes advantageous for longer time series, due to its 
linear sampling complexity ($M \propto N$), which scales more favourably 
than the $\mathcal{O}(N^2)$ complexity of RQA\_woRP,
i.e., the computational savings grow with data length.

The results presented here demonstrate typical behaviour for a 
continuous chaotic system (R\"ossler attractor). Different dynamical 
regimes may affect the performance of these approaches differently. 
For instance, discrete maps like the logistic map generate RPs 
with shorter, more heterogeneous line structures. This leads to 
faster sampling (as valid line starts are found more frequently) 
but potentially larger uncertainties, particularly when $M$ is 
insufficient to adequately represent the line length distribution 
(Fig.~\ref{fig_logistic_periodic}A, C).
In contrast, periodic systems have more homogeneous RPs with very long,
non-interrupted, 
but less frequent diagonal lines. This reduces the probability of 
finding lines through random sampling, increasing computation time
(can be even worse than the standard RQA\_RP approach). 
However, the limited diversity of line lengths ensures accurate 
results even for small $M$ (Fig.~\ref{fig_logistic_periodic}B, D).

Sampling approaches are most efficient for homogeneously structured 
RPs, characteristic of stationary dynamics. Non-stationary systems 
with heterogeneous RPs are expected to lead to larger uncertainties in RQA\_Samp results, 
requiring either the use of RQA\_woRP or increasing $M$. Systematic 
guidelines for selecting $M$ in non-stationary cases warrant future 
investigation.

The presented approaches also come with potential trade-offs. The 
sampling of structures introduces an additional source of variability, 
and its influence on the robustness and reliability of different RQA 
measures still needs to be systematically evaluated. Future studies 
should investigate how sampling density, noise, and parameter settings 
affect the stability of the obtained measures, particularly for 
non-stationary dynamics where line structures may be heterogeneously 
distributed. Another important aspect concerns the integration of 
established correction schemes, such as those for removing sojourn 
points \cite{gao99,kraemer2019,marwan2007} or mitigating border 
effects \cite{censi2004,kraemer2019}, into the sampling framework.

Beyond methodological refinements, several directions could further 
enhance the computational efficiency and applicability of these 
approaches. Combining the proposed sampling strategy with parallel 
computing architectures could improve scalability for very large 
datasets and real-time applications. Additionally, the current 
implementations could benefit from further 
refinements. Replacing the fixed sampling size $M$ with an adaptive 
convergence scheme -- which stops sampling once results stabilise within 
a predefined threshold -- could optimise the trade-off between accuracy 
and computation time. Combined with low-level compiler optimisations 
such as JIT compiler hints and vectorisation \cite{sherrington2024}, 
these enhancements could yield additional modest speedups.

The strategies discussed here are not limited to the diagonal line 
measures. They can be analogously applied to RQA measures based on 
vertical line structures \cite{marwan2002herz} and to recurrence 
network measures, such as clustering coefficient and transitivity 
\cite{marwan2009b}, broadening the scope of energy-efficient 
recurrence analysis.

In summary, the combination of direct line detection and sampling represents a 
promising way forward in the efficient computation of RQA and related measures. 
Beyond the methodological advantages, the reduction in computational cost also 
contributes to the principles of green computing by lowering energy consumption, 
which makes the approach more sustainable for large-scale or long-term data analysis.

\section*{Acknowledgements}

The author thanks Felipe Eduardo Lopes da Cruz, Thiago de Lima Prado, 
and Sergio Roberto Lopes for inspiring discussions on sampling 
microstates and for motivating this research.
The author gratefully acknowledges support from the German 
Academic Exchange Service (DAAD)
and Coordenação de Aperfeiçoamento de Pessoal de Nível Superior (CAPES)
via the project ``Recurrence quantifiers as features for 
machine-learning-decision-making processes'' (Grant No.~57705568), 
as well as from the Ministry of Research, Science and 
Culture (MWFK) of Land Brandenburg for supporting this project by providing 
resources on the high-performance computer system ``Eunice Newton Foote''
at the Potsdam Institute for Climate Impact Research (Grant No.~22-Z105-05/002/001).
For the RQA estimations using the microstates approach, the recently
published Julia package \texttt{RecurrenceMicrostatesAnalysis.jl} \cite{ferreira2025}
has been used.

\section*{Data and code availability}

Code used for this analysis and to reproduce the results and figures
of this study are available at Zenodo: \url{https://zenodo.org/10.5281/zenodo.17620044}.

\section*{Appendix}

\renewcommand{\thefigure}{A\arabic{figure}}
\setcounter{figure}{0}



\subsection*{Energy consumption per run}

Assuming a constant power draw $P$ (e.g., $P = \SI{50}{\watt}$) and a computation time $t$
measured in seconds, the energy consumption per run is
$$
E = P\frac{t}{3600} \ [\unit{\Wh}].
$$
Accordingly, the energy saving obtained when replacing a reference method with 
runtime $t_\text{ref}$ by a faster method with runtime $t$ is
$$
\Delta E = P \frac{t_\text{ref} - t}{3600}\ [\unit{\Wh}].
$$

\subsection*{Performance of the algorithm and further dynamical systems}

\begin{figure}[htbp]
\begin{center}
   \includegraphics[width=.8\columnwidth]{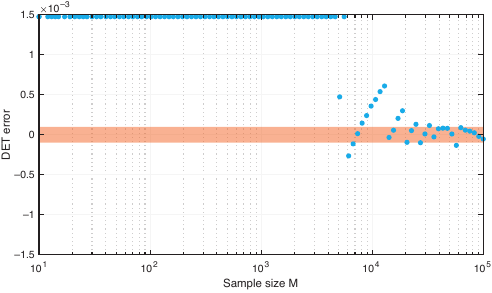}
   \caption{Estimation error of DET derived from recurrence microstates
   approximation for the R\"ossler system with 25,000 data points
   and sampling size ranging from $M=10$ to $100,000$. 
   The orange region indicates the error range of $\pm 10^{-4}$.
   To achieve estimation errors below $10^{-4}$, the sampling size must be 
   at least $40,000$. Up to $M=5,000$, the
   estimated DET remains at $1.00$, resulting in the constant deviation.
   }\label{fig_microstates_samplesize}
\end{center}
\end{figure}

\begin{figure}[htbp]
\begin{center}
   \includegraphics[width=.8\columnwidth]{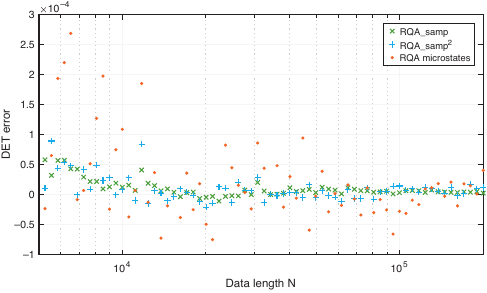}
   \caption{Estimation error $DET_\text{sampled} - DET_\text{true}$ for DET estimations using RQA\_Samp, 
   RQA\_Samp$^2$, and recurrence microstates
   approximation for the R\"ossler system for increasing data length $N$.
   The sampling was $M=100,000$ for RQA\_Samp and microstates RQA,
   and $M=5,000$ for RQA\_Samp$^2$. Even for a very small sampling number,
   the RQA\_Samp approach has small estimation errors $<10^{-4}$.
   }\label{fig_datasize_error}
\end{center}
\end{figure}

\begin{figure}[htbp]
\begin{center}
   \includegraphics[width=\columnwidth]{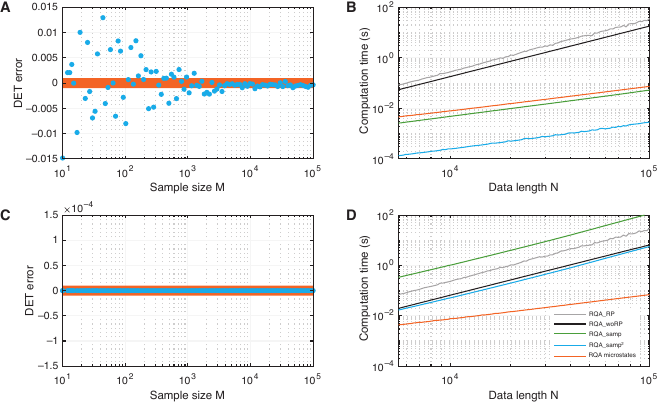}
   \caption{(A, C) Estimation error $DET_\text{sampled} - DET_\text{true}$ and 
   (B, D) computation
   times as in Figs.~\ref{fig_samplesize} and \ref{fig_datasize} for the
   (A, B) the logistic map and (C, D) a periodic signal.
   In (A, C) results for $N = 25,000$ and using RQA\_Samp$^2$ are shown. 
   }\label{fig_logistic_periodic}
\end{center}
\end{figure}

\clearpage
\subsection*{Recommendation for required data length}

While a more systematic study is necessary to get good recommendations
on required time series lengths for the proposed RQA\_samp$^2$ approach,
we can already gather some clues.

\begin{figure}[htbp]
\begin{center}
   \includegraphics[width=\columnwidth]{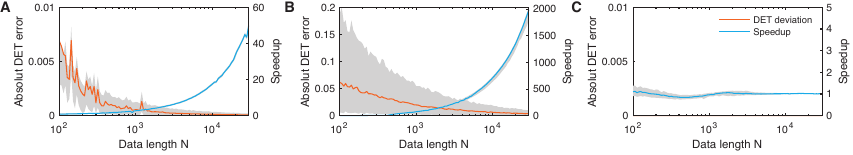}
   \caption{Speedup of computation time ($t_{\text{RQA\_samp}^2}/t_\text{RQA\_woRP}$) and absolute estimation error 
   $|DET_\text{sampled} - DET_\text{true}|$ as function of data length
   using the RQA\_samp$^2$ approach
   for (A) the R\"ossler system, (B) the logistic map, and (C) a periodic signal.
   Shaded areas indicate the 5\% and 95\% quantiles obtained from 500 repetitions;
   solid lines denote the median. 
   }\label{fig_criterion_datasize}
\end{center}
\end{figure}

Errors in estimating the RQA measure and computation times with RQA\_samp$^2$
depend on the characteristics of the data. 
Data that cause relatively homogeneously distributed long lines in an RP, such as for
R\"ossler or Lorenz systems, show estimation errors that are already quite small for short data 
lengths, with $100<N<1,000$. The speedup in computation time becomes significant 
for $N>1,000$ (Fig.~\ref{fig_criterion_datasize}A).
For systems that generate many short lines and single points, as typical for maps,
the errors are generally larger, but the computation time benefits already
significantly for small $N$ (Fig.~\ref{fig_criterion_datasize}B). To achieve 
reasonable results, at least $N>2,000$ are necessary, although this number 
can be lowered by increasing the sampling size $M$.
In contrast, sparse RPs with consistent patterns cause numerous unsuccessful 
sampling tries, leading to limited speedup but very slight errors
(Fig.~\ref{fig_criterion_datasize}C). Therefore, this approach might be 
less effective for periodic systems compared to signals with more complex 
or irregular dynamics.

\clearpage

\bibliographystyle{plainnat}
\bibliography{rp,misc,mybibs}

\end{document}